\documentclass[[preprint,12pt
  ]
  {revtex4}
  
 \usepackage{amsfonts}
\usepackage{amsmath}
\usepackage{amssymb}
\usepackage{graphicx}
\usepackage[usenames]{color}

\begin{document}

\title{Non-Newtonian Properties of Relativistic Fluids}


\author{Tomoi Koide}
\affiliation{
{Frankfurt Institute for Advanced Studies, 
D-60438 Frankfurt am Main, Germany}
}

\begin{abstract}
We show that relativistic fluids behave as non-Newtonian fluids.
First, we discuss the problem of acausal propagation in the diffusion equation and 
introduce the modified Maxwell-Cattaneo-Vernotte (MCV) equation.
By using the modified MCV equation, we obtain the causal dissipative relativistic (CDR) 
fluid dynamics, where unphysical propagation with infinite velocity does not exist.
We further show that the problems of the violation of causality and instability are 
intimately related, and the relativistic Navier-Stokes equation is inadequate as the 
theory of relativistic fluids.
Finally, the new microscopic formula to calculate the transport coefficients of 
the CDR fluid dynamics is discussed.
The result of the microscopic formula is consistent with that of the Boltzmann equation, i.e., Grad's moment method.
\end{abstract}

\maketitle


\section{Introduction}

\begin{figure}
  \includegraphics[height=.25\textheight]{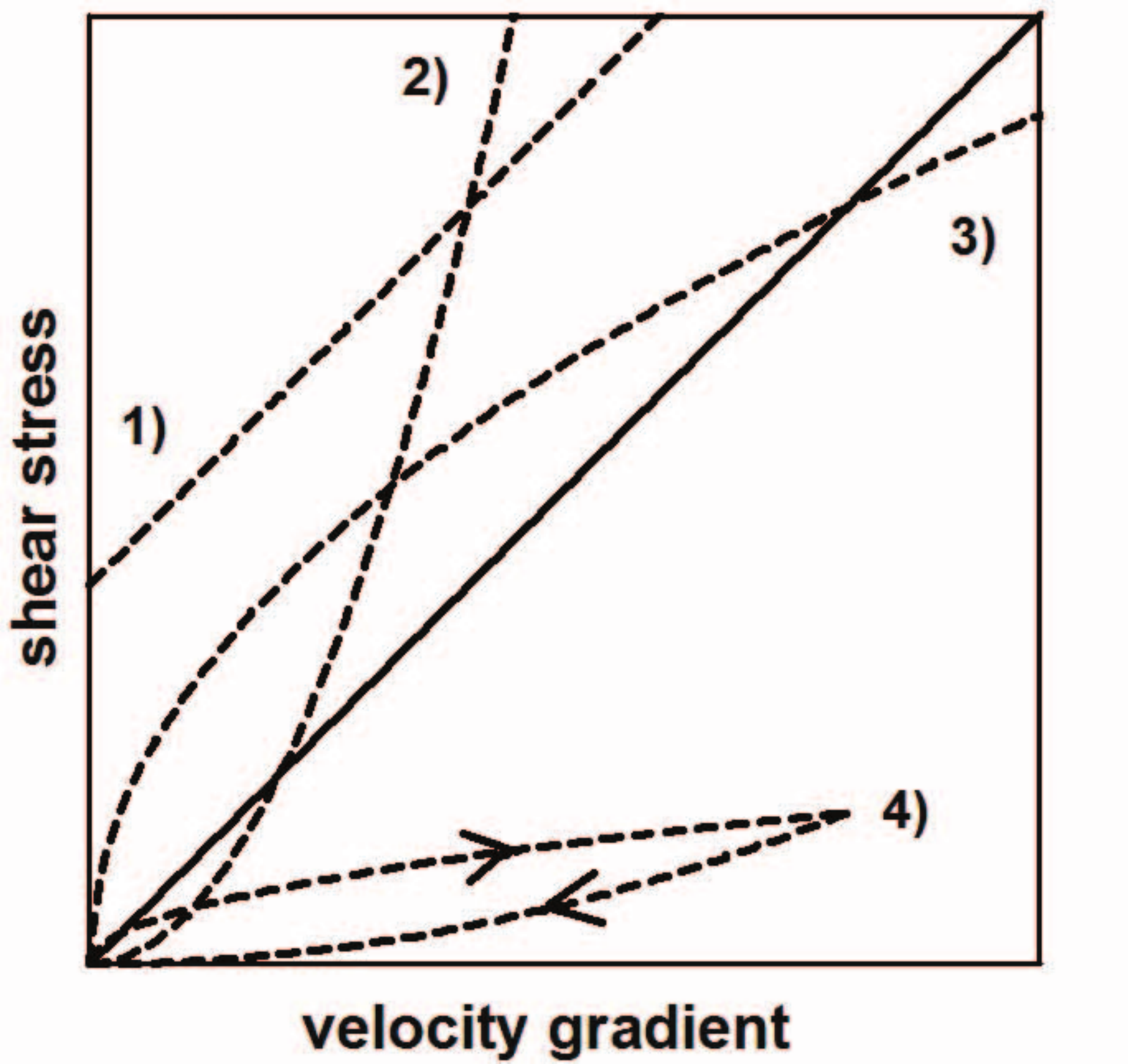}
  \caption{Shear stresses of Newtonian and non-Newtonian fluids as a function of the velocity gradient. The solid line denote a Newtonian fluid. The dashed lines represents 1) Bingham flow, 2) dilatant fluids, 2) pseudoplastic and 4) thixotropic fluids, respectively.  }
  \label{fig1}
\end{figure}

Typical examples of fluid are water and air, whose dynamics is 
described by the Navier-Stokes (NS) equation. 
These are called Newtonian fluids.
There are, however, various fluids called non-Newtonian fluids, which cannot be described by 
the NS equation.
The difference between these two types of fluid comes from the 
behavior of the shear stress tensor. 
In Fig. \ref{fig1}, the various shear stress tensors are shown as a 
function of the gradient of the fluid velocity. 
When the shear stress tensor increases proportionally with the velocity gradient, 
the fluid is Newtonian, which is denoted by the solid line. 
Non-Newtonian fluids exhibit a more complex behavior as is shown by the dashed lines.
The Bingham flow 1) represents a similar linear relation but the shear stress tensor does not 
disappear even in the vanishing velocity-gradient limit. 
The dilatant fluid 2) and pseudoplastic 3) show non-linear dependences.
The shear stress tensor of the thixotropic fluids 4) depends on time.

If the dynamics of relativistic many-body systems can be described by using 
coarse-grained equations such as fluid dynamics, 
is the behavior of relativistic fluids Newtonian or non-Newtonian ?
In order to answer this question, we will start our discussion from diffusion processes, 
because the problem which we will encounter in relativistic fluid dynamics has already 
appeared in the diffusion equation.

\section{diffusion equation and MCV equation}

We consider a random walk process, where 
a particle moves to left or right on a one-dimensional lattice with 
equal probability. 
The probability distribution function $P(x,t)$ satisfies 
$P(x,t+dt) = (P(x-dx,t) + P(x+dx,t))/2$, 
where $dx$ and $dt$ are the size of lattice and time steps, respectively. 
In the continuum limit, we obtain the diffusion equation, 
\begin{equation}
\partial t P(x,t) = D \partial^2_x P(x,t),
\end{equation}
with the definition of the diffusion coefficient, 
$D = \lim_{dt,dx \rightarrow 0} (dx)^2/dt$.
Then the particle can move by $dx$ at each time step 
$dt$ and the velocity of the particle 
is given by $v=dx/dt$. Note that the continuum limit should be taken by fixing 
$D$. This leads to the infinite velocity,
\begin{equation}
\lim_{dt,dx \rightarrow 0} v = \lim_{dt,dx \rightarrow 0} \sqrt{D/dt} = \infty .
\end{equation}

Let us consider a possible modification of the diffusion equation 
to avoid this violation of causality. Remember that the diffusion equation consists of 
two structures. One is the equation of continuity, 
\begin{equation}
\partial_t n + \nabla {\bf J} = 0, \label{eoc}
\end{equation}
where $n$ is a conserved density and ${\bf J}$ is a current.
The other is the definition of ${\bf J}$. To obtain the diffusion equation, 
we assume that ${\bf J}$ is proportional to  
the corresponding thermodynamic force ${\bf F}$, 
\begin{equation}
{\bf J} = -D {\bf F} = -D \nabla n . \label{fick}
\end{equation} 
Here ${\bf F} = \nabla n$ for the diffusion process. 
This is called Fick's law.

The equation of continuity (\ref{eoc}) should be satisfied for any conserved density. 
Thus, if it is possible to derive a modified diffusion equation consistent with causality, 
only Eq. (\ref{fick}) can be changed.
As a matter of fact, from a microscopic theory such as the linear response theory, 
a more general expression of 
${\bf J}$ is given by the time convolution integral, 
\begin{equation}
{\bf J}(t) = -\int^t ds G(t-s) {\bf F}(s), \label{general_jf}
\end{equation}
where $G(t)$ is the memory function which is given by the time correlation function of 
microscopic degrees of freedom. 
Thus the time scale of the memory function is characterized by the microscopic time scale.
If the time scale of macroscopic variables such as ${\bf J}$ and ${\bf F}$ is clearly separated 
from the microscopic one, we can approximately replace the time dependence of $G(t)$ with the 
Dirac delta function, $G(t) = D \delta (t)$ and then we can reproduce Fick's law (\ref{fick}).

When, however, the time scales 
are not clearly separated, the time dependence of $G(t)$ should be taken into account.
As a simplest choice, we use the exponential form,
\begin{equation}
G(t) = \frac{D}{\tau_R} e^{-t/\tau_R} ,
\end{equation}
where $\tau_R$ is the relaxation time which characterizes the microscopic time scale.
Substituting into Eq. (\ref{general_jf}) and operating the time derivative, 
we obtain 
\begin{equation}
\tau_R \partial_t {\bf J} (t)+ {\bf J} (t)= -D{\bf F} (t). \label{mcv}
\end{equation}
This is the so-called Maxwell-Cattaneo-Vernotte (MCV) equation.
When there is a clear separation of microscopic and macroscopic time scales, $\tau_R$ vanishes and 
the MCV equation is reduced to Fick's law (\ref{fick}).

\begin{figure}
  \includegraphics[height=.25\textheight]{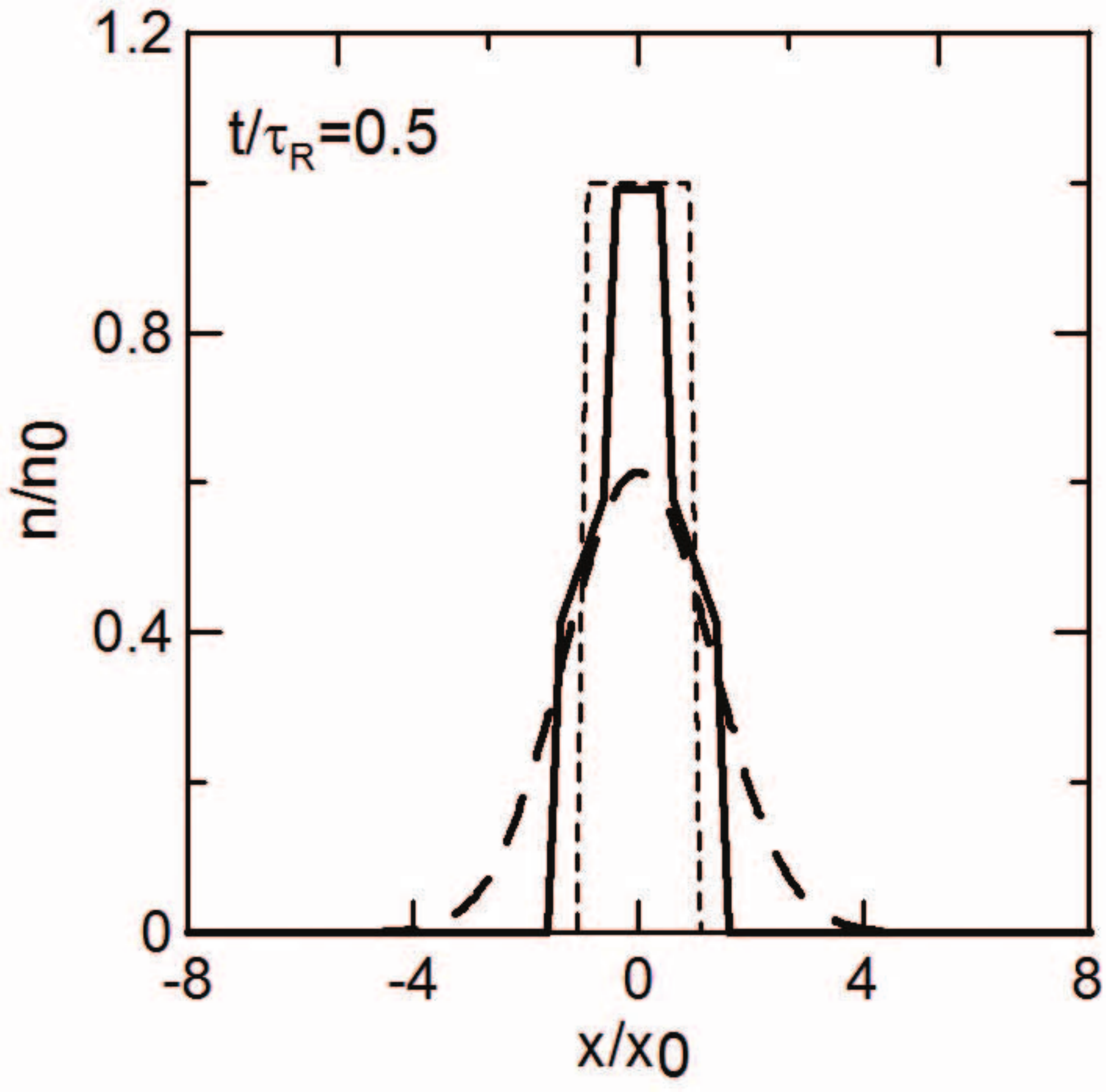}
  \includegraphics[height=.25\textheight]{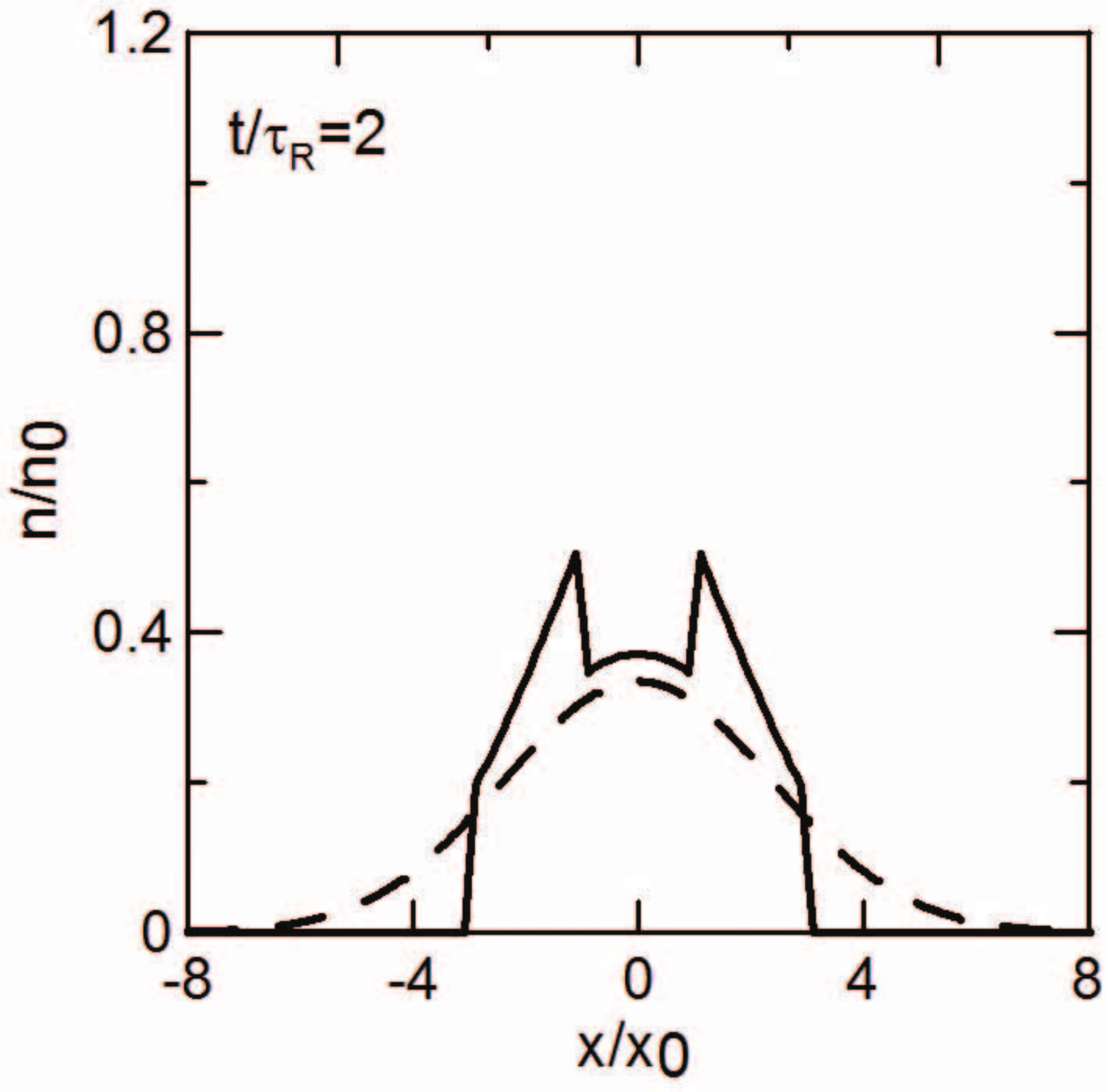}
  \includegraphics[height=.25\textheight]{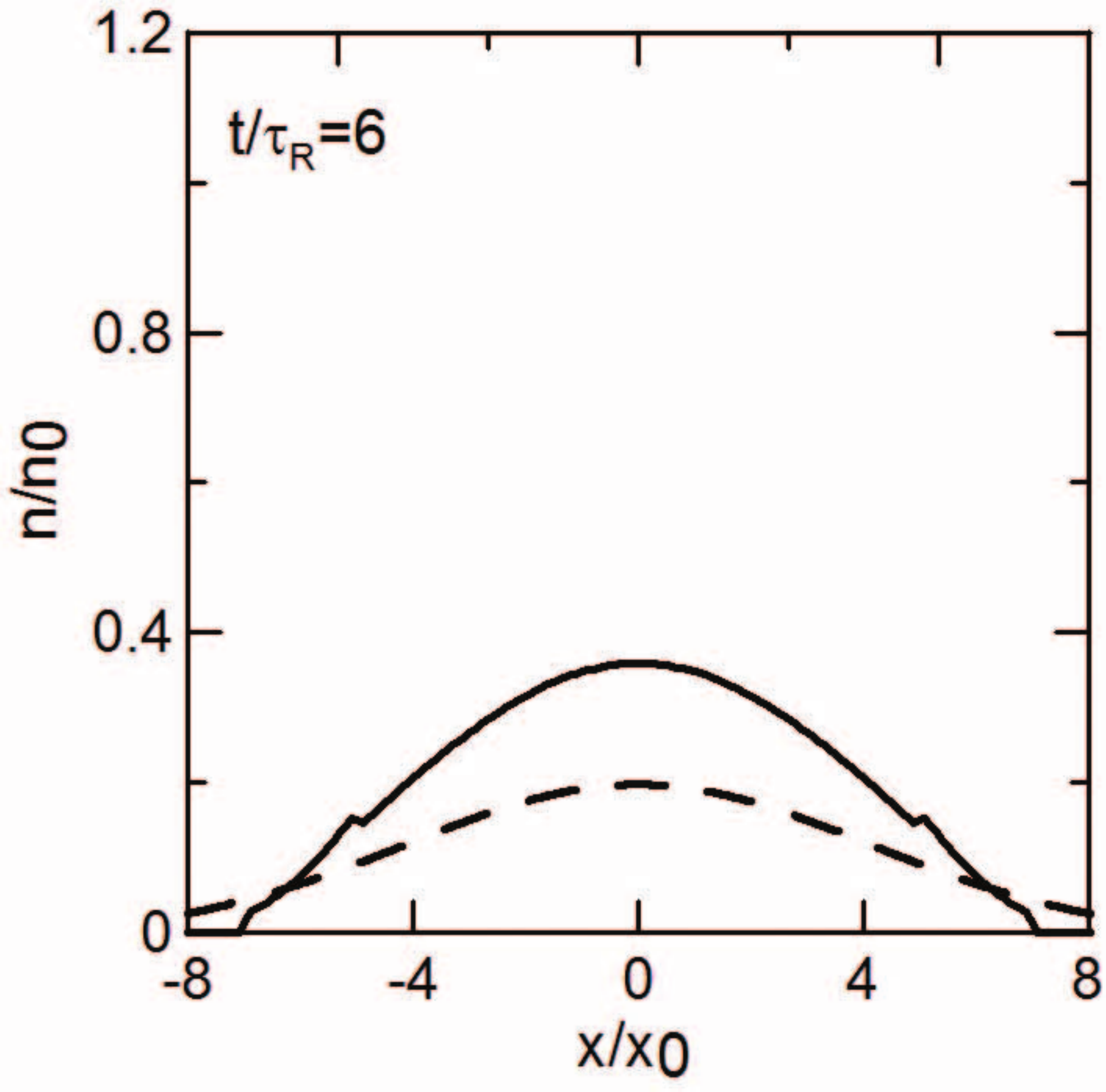}
  \caption{The time evolutions of the diffusion equation (dashed line) and the MCV equation (solid line) 
for $t/\tau_R = 0.5$, $2$ and $6$. The initial condition is shown by the dotted line at $t/\tau_R = 0.5$.}
  \label{fig2}
\end{figure}

By eliminating {\bf J} from the equation of continuity (\ref{eoc}) with
the MCV equation (\ref{mcv}), we have the telegraph equation for $n$, 
\begin{equation}
\tau_R \partial^2_t n + \partial_t n = D \nabla^2 n . \label{tele}
\end{equation}
The analytic solution is given in Ref. \cite{mf}.
For example, the solution of 1+1 dimensional system is given by 
\begin{eqnarray}
n(x,t) 
&=& \frac{e^{-t/(2\tau_R)}}{2}\{ n_0 (x+vt) + n_0 (x-vt) \}\nonumber \\
&& + e^{-t/(2\tau_R)} \int^{x+vt}_{x-vt} dx' 
\left\{ \frac{1}{4v\tau_R} n_0 (x') + \frac{1}{v}\partial_x J_0(x') \right\}
I_0 \left( \sqrt{v^2 t^2 - (x-x')^2}/(2v\tau_R)\right) \nonumber \\
&& + e^{-t/(2\tau_R)} \int^{x+vt}_{x-vt} dx'
\frac{1}{2v\tau_R} \partial_t 
I_0 \left( \sqrt{v^2 t^2 - (x-x')^2}/(2v\tau_R)\right),
\end{eqnarray} 
where $I_0$ is the modified Bessel function, and $n_0$ and $J_0$ are 
initial conditions of the conserved density and corresponding current, respectively.
The maximum propagation speed of this equation is characterized by $v$, 
\begin{equation}
v = \sqrt{D/\tau_R},
\end{equation}
which diverges in the diffusion limit ($\tau_R = 0$).
In Fig. \ref{fig2}, the time evolutions of the diffusion equation and 
the telegraph equation are shown by the dashed and solid lines. 
We use $D = \tau_R = 1$, leading to $v=1$. The initial denstiy distribution $n_0(x)$ is plotted 
in the figure of $t/\tau_R = 0.5$ by the dotted line, and we further set $J_0 (x)=0$. 
One can see that, because of the memory effect, there exists a non-trivial structure 
at the boundary of the expanding $n$, reflecting the initial distribution 
in the telegraph equation, 
although the diffusion equation always shows the Gaussian forms.
Of course, because of the finite propagation speed, the expansion of 
the telegraph equation is slower than that of the diffusion equation.
See also Ref. \cite{weiss}.

In table \ref{tab:a}, the comparison of the diffusion equation and the MCV equation 
is summarized.
If the diffusion equation is a coarse-grained dynamics of the underlying microscopic physics, 
it should be derived from a microscopic theory using systematic coarse-graining.
As a matter of fact, as is discussed in textbooks, it is believed that 
the diffusion equation can be derived with the projection operator method. 
However, we should note that one non-trivial approximation is used in this 
derivation. 
As a matter of fact, 
it was recently found that, when such non-trivial approximation is not applied, 
the MCV equation is obtained instead of the diffusion equation \cite{koide_diff}.

Correspondingly, the microscopic Hamiltonian which describes a diffusion process 
should have a symmetry associated with the conserved density, 
and we can derive the corresponding sum rule.  
The sum rule determines the initial time evolution of the conserved density.
The telegraph equation is consistent with this sum rule, although the diffusion equation is not 
\cite{koide_diff,kadanoff}.

As is well known in linear irreversible thermodynamics, 
the positivity of the entropy production is algebraically satisfied when there is a 
simple linear relation between ${\bf J}$ and ${\bf F}$. 
However, this linear relation is not satisfied in the MCV equation.
Moreover, when there is no clear separation of time scales, 
we cannot assume quasi-adiabatic changes of thermodynamical variables, and then 
heat will play a more fundamental role instead of entropy. 
Thus the second law for the MCV equation is not trivial.
As for the problem of positivity, see Ref. \cite{kampen}.
\begin{center}
\begin{table}
\begin{tabular}{c c c}
\hline
                  & diffusion & MCV \\
\hline
propagation speed & $\infty$ & $\sqrt{D/\tau_R}$ \\
microscopic derivation & ? & \textcircled{\tiny} \\
sum rule & $\times$ & \textcircled{\tiny} \\
2nd law of thermodynamics & \textcircled{\tiny} & ? \\
positivity & \textcircled{\tiny} & ? \\
\hline
\end{tabular}
\caption{Comparison of the diffusion equation and the MCV equation}
\label{tab:a}
\end{table}
\end{center}

\section{modified MCV equation}

So far, we have considered currents induced by gradients of $n$.
However, when there exists a macroscopic velocity ${\bf v}$, 
the total current is given by two contributions, 
\begin{equation}
{\bf J}_{tot} = n {\bf v} + {\bf J},
\end{equation}
and the conserved density $n$ should satisfy the equation of continuity with ${\bf J}_{tot}$ instead of ${\bf J}$ itself.
In this more general case, the MCV equation is modified \cite{dkkm4}.

Remember that the conserved density $n$, the velocity ${\bf v}$ 
and the thermodynamics force $\nabla n$ are defined by the 
averaged quantities of particles which are contained inside a small but finite 
volume $V^*$, commonly refered to as fluid cell \cite{wey}.
Thus, Fick's law should be expressed for quantities per unit cell,
\begin{equation}
[{\bf J}(t)V^*(t)] = -D [(\nabla n(t)) V^*(t)].
\end{equation}
Note that $V^*(t)$ is a function of time because of the deformation induced by the macroscopic flow.
In the case of Fick's law, however, $V^*(t)$ is a common factor and we can finally 
reproduce the ordinary result (\ref{fick}).
On the other hand, Eq. (\ref{general_jf}) is modified as 
\begin{equation}
{\bf J}(t)V^*(t)  = -\int^t ds G(t-s) {\bf F}(s)V^*(s) . \label{general_jf2}
\end{equation}
Because of the different time dependence of $V^*$, the deformation of the fluid cell affects 
the evolution of the current ${\bf J}$.
The dynamics of the fluid cell is determined from a geometrical argument.
Now we introduce three vectors to define the volume of the fluid cell, 
$V^* = (\vec{\xi}_1 \times \vec{\xi}_2)\times \vec{\xi}_3$. 
Because of the macroscopic velocity ${\bf v}$, any vector is shifted to 
${\bf r} \rightarrow {\bf r}' = {\bf r} + {\bf v} dt $. 
Finally we can derive the following equation for the deformation of the fluid cell,
\begin{equation}
\frac{d}{dt} V^* (t) 
\equiv \left( \partial_t + {\bf v}(t)\cdot \nabla \right) V^*(t) = V^{*}(t)\nabla \cdot {\bf v}(t) .
\label{eq_v*}
\end{equation}
Combining Eqs. (\ref{general_jf2}) and (\ref{eq_v*}), we obtain the modified MCV equation 
\cite{dkkm4}, 
\begin{equation}
\tau_R \frac{d}{dt} {\bf J}(t) + \tau_R {\bf J}(t) (\nabla \cdot {\bf v}(t)) 
+ {\bf J}(t) = -D {\bf F}(t) .
\end{equation}
Interestingly enough, the final result does not depend on the volume of the fluid cell $V^*$.

The second term on the l.h.s. is a new term which does not exist in the MCV equation, 
and disappears when there is no macroscopic velocity. 
In the vanishing $\tau_R$ limit, this equation still reproduces Fick's law (\ref{fick}).

The importance of the dynamics of volume elements is discussed also by Brenner \cite{brenner},
where an additional velocity variable is introduced.
In our case, however, we do not introduce any additional velocity.

So far, we have discussed the modification of the diffusion equation by introducing a memory effect.
There are, however, several different approaches to obtain modified diffusion equations.
For example, there are attempts to obtain the MCV equation from 
the random walk model including memory effect, which is called the persistent random walk. 
So far, this approach reproduces the MCV equation 
only for the 1+1 dimensional case \cite{goldstein}.
In our work, the memory function $G$ has been assumed to have the exponential form. 
Other possibilities for the memory function are summarized in Ref. \cite{joseph}.
Moreover, as is discussed in Ref. \cite{kath}, the problem of acausality may be solved by 
introducing a non-linear effect.
van Kampen also discussed modifications of the diffusion 
equation, associated with the heat conduction of a photon gas \cite{kampen}.
The memory effect for phase separation is discussed in Ref. \cite{kkr}

\section{relativistic fluid dynamics}

The problem of the violation of causality becomes more important in deriving 
relativistic fluid dynamics.
In the following, we use the natural unit, where $c=\hbar = 1$, and the metric 
$g^{\mu\nu} = {\it diag}\{1,-1,-1,-1\}$.

In the derivation, we first choose gross variables which are necessary to 
extract the macroscopic motion of many-body systems.
If the chosen variables are not enough, the derived fluid dynamics will show 
unphysical behaviors, such as instability \cite{dkkm3,pu} and the divergent transport 
coefficients \cite{koide_diff}. 
Unfortunately, there is no systematic procedure to collect 
the complete set of gross variables \cite{kawasaki}, but 
normally, conserved densities are chosen.
In discussing phase transitions, order parameters are also one of the candidates of 
gross variables.
Here we do not consider phase transitions and conserved charges.
Then the gross variables are the energy density $\varepsilon$ 
and fluid velocity $u^{\mu}$ contained in the conserved energy-momentum tensor.

In the idealized case, the 
the energy-momentum tensor $T^{\mu\nu}$ 
is a function only of $\varepsilon$ and $u^{\mu}$.
Then, by applying a Lorentz transformation and using the 
definition of the energy density and pressure $P$, we obtain 
$T^{\mu\nu} = (\varepsilon + P)u^\mu u^\nu - g^{\mu\nu} P$. 
Note that $P$ is calculated by the equation of state.
Since $T^{\mu\nu}$ is conserved, we have 
\begin{equation}
\partial_\mu T^{\mu\nu} = 0.
\end{equation}
This is the relativistic Euler equation.

However, in general, $T^{\mu\nu}$ cannot be expressed only by $\varepsilon$ and $u^{\mu}$.
We represent this additional component by another second rank tensor $\Pi^{\mu\nu}$. 
The most general $T^{\mu\nu}$ is, then, given by 
$ T^{\mu\nu} = (\varepsilon + P)u^\mu u^\nu - g^{\mu\nu} P +\Pi^{\mu\nu}$.
Conventionally, $\Pi^{\mu\nu}$ is expressed using the trace part $\Pi$ and traceless part 
$\pi^{\mu\nu}$ as $\Pi^{\mu\nu} = \pi^{\mu\nu} - (g^{\mu\nu} -u^\mu u^\nu) \Pi$ 
\footnote{The heat conduction is neglected, because it finally disappears by using the 
definition of the fluid velocity.}.
Finally $T^{\mu\nu}$ is expressed as
\begin{equation}
T^{\mu\nu} = (\varepsilon + P +\Pi)u^\mu u^\nu - g^{\mu\nu} (P+\Pi) + \pi^{\mu\nu},
\end{equation}
and $\Pi$ and $\pi^{\mu\nu}$ are the bulk viscous pressure and the shear stress tensor, 
respectively, satisfying the orthogonality condition $u_\mu \pi^{\mu\nu} = 0$.

The determination of these viscous terms is our next task. 
In the traditional Landau-Lifshitz theory \cite{ll}, these are induced instantaneously by 
the corresponding thermodynamic force, similarly to the diffusion equation, 
\begin{eqnarray}
\Pi = -\zeta \theta ~~~~~~~~~~~~
\pi^{\mu\nu} = 2 \eta \sigma^{\mu\nu}, 
\end{eqnarray} 
where $\zeta$ and $\eta$ are the bulk and shear viscosities, respectively.  
The thermodynamic forces $\theta$ and $\sigma^{\mu\nu}$ 
are defined by 
\begin{eqnarray}
\theta &=& \partial_\mu u^\mu , \\
\sigma^{\mu\nu} 
&=& \frac{1}{2}\left( \partial^\mu u^\nu + \partial^{\nu} u^\mu 
- \frac{2}{3} (g^{\mu\nu} - u^\mu u^\nu) \theta \right) 
\equiv \Delta^{\mu\nu\lambda \delta} \partial_\lambda u_\delta .
\end{eqnarray}
When we use these definitions of the viscous terms, we obtain 
the relativistic NS equation. 
Because of the instantaneous production of the viscous terms, 
this equation contains the propagation with infinite speed, 
as is shown later.

As was discussed for diffusion processes, 
the violation of causality is solved by introducing the memory effect previously discussed.
Because of the existence of the fluid velocity $u^\mu$, the modified MCV equation 
should be applied. The relativistic representation of 
the modified MCV equation is 
\begin{equation}
\tau_R u^\mu \partial_\mu {\bf J} + \tau_R {\bf J} \theta = -D {\bf F}.
\end{equation}
Thus the viscous terms satisfying causality are 
given by 
\begin{eqnarray}
\tau_\Pi u^\mu \partial_\mu \Pi + \tau_\Pi \Pi \theta + \Pi 
&=& - \zeta \theta,  \label{eq_Pi} \\
\tau_\pi \Delta^{\mu\nu\lambda\delta} u^\alpha \partial_\alpha \pi_{\lambda \delta} 
+ \tau_\pi \pi^{\mu\nu} \theta + \pi^{\mu\nu} 
&=& 2\eta \sigma^{\mu\nu}, \label{eq_pi}
\end{eqnarray}
where $\tau_\Pi$ and $\tau_\pi$ are the relaxation times of $\Pi$ and $\pi^{\mu\nu}$, 
respectively.
Here the projection operator $\Delta^{\mu\nu\lambda\delta}$ 
is necessary to satisfy the orthogonality relation.
In the following, we call this theory the causal dissipative relativistic (CDR) fluid dynamics.

The second terms on the l.h.s. of Eqs. (\ref{eq_Pi}) and (\ref{eq_pi}) comes from the 
deformation of the fluid cells.
As is shown in Figs. 3--8 of Ref. \cite{dkkm4}, these terms are necessary to implement stable numerical 
calculations with ultra-relativistic initial conditions.

Here, we simply assume that the thermodynamic forces of the modified MCV equation are the same as 
those of the relativistic NS equation.
However, it may be possible to consider the higher order corrections to 
the thermodynamic forces, as is discussed in the Burnett equation \cite{garcia2}.

\section{causality and stability of relativistic fluids}

When relativistic fluids are described by the relativistic NS equation, 
the fluid is Newtonian because there is a proportional relation between 
$\pi^{\mu\nu}$ and $\sigma^{\mu\nu}$.
On the other hand, the CDR fluid dynamics describes relativistic non-Newtonian fluids because $\pi^{\mu\nu}$ is determined by solving the differential equation.

\begin{figure}
  \includegraphics[height=.25\textheight]{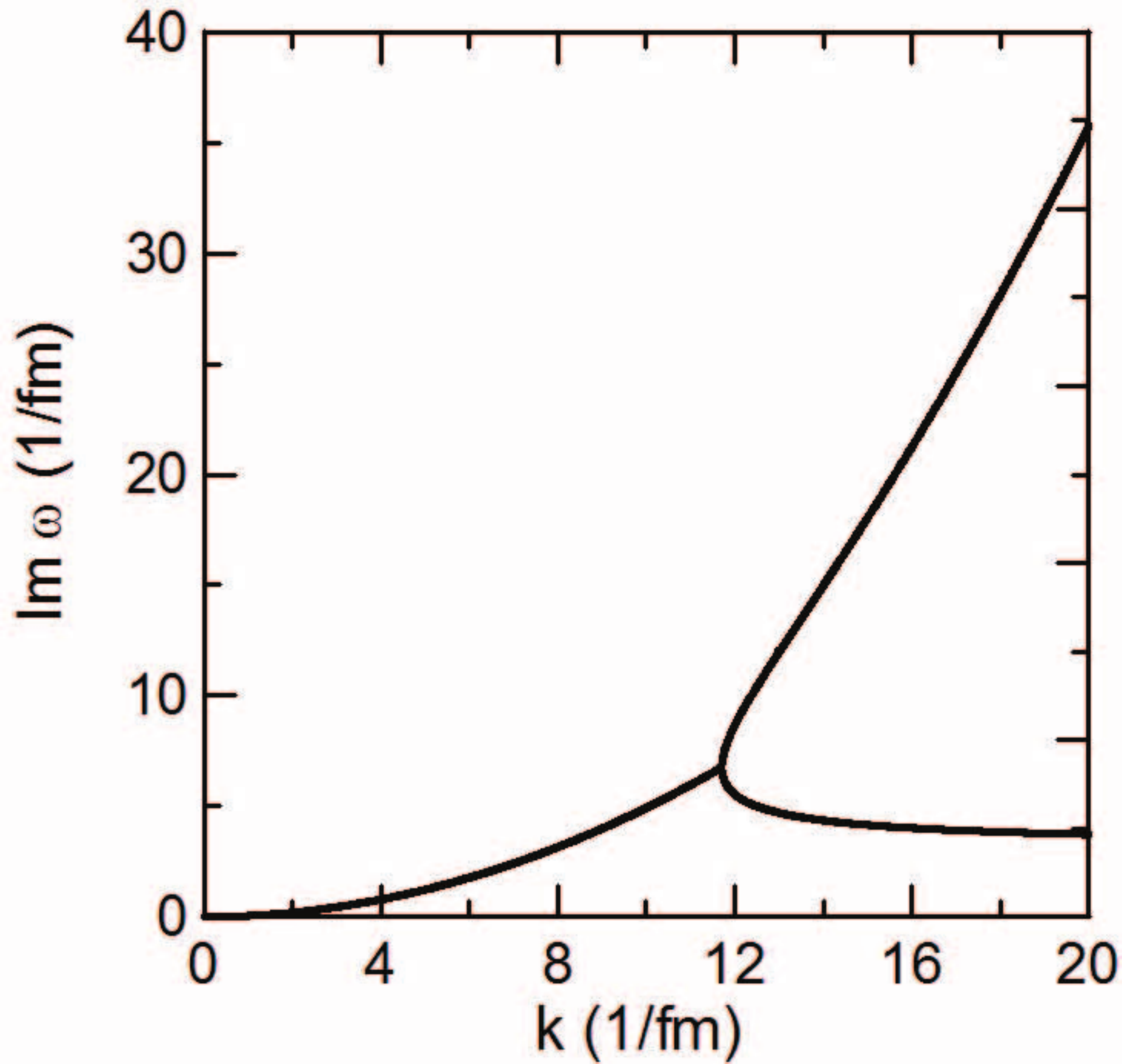}
  \includegraphics[height=.25\textheight]{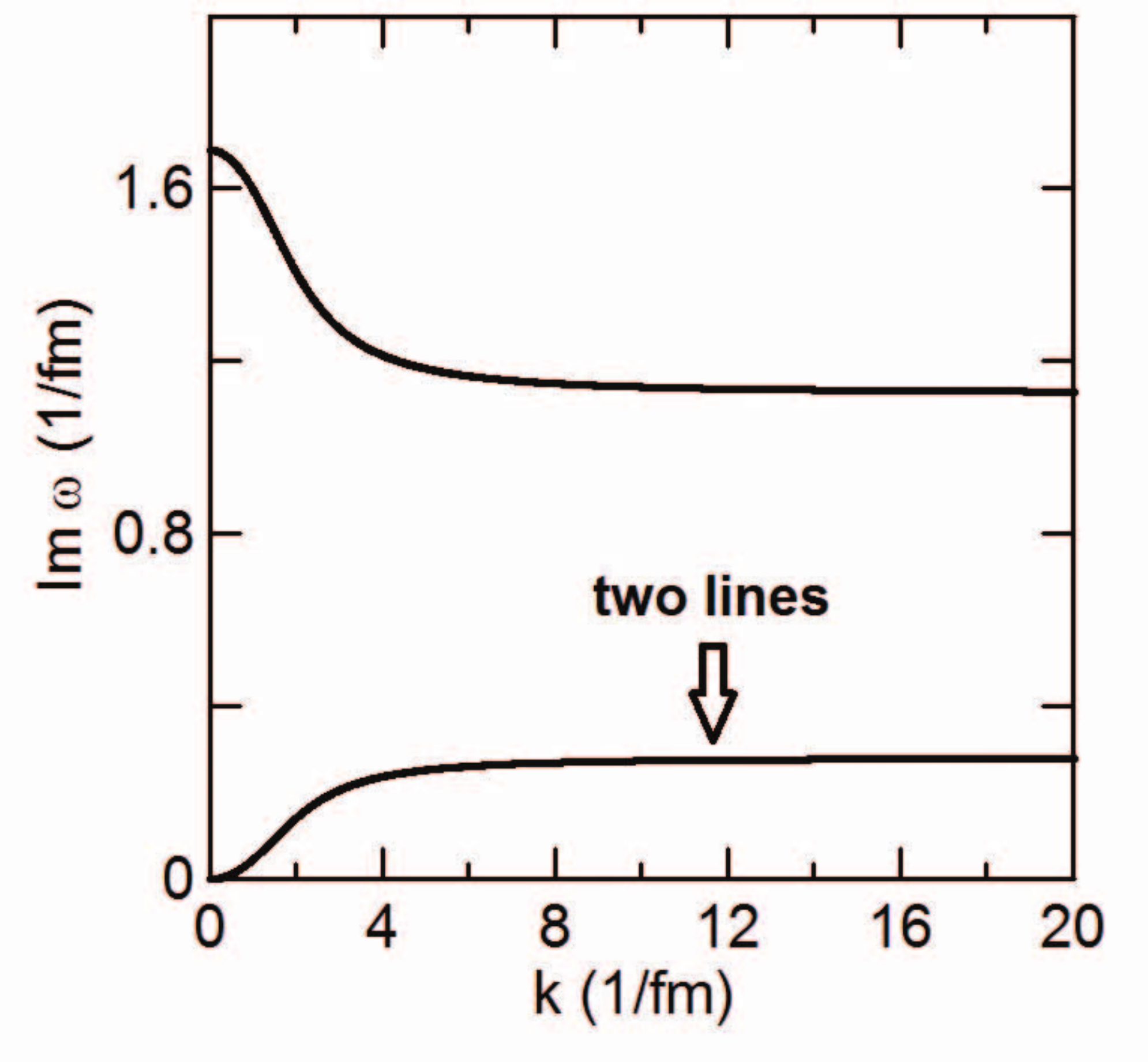}
  \caption{The imaginary part of the dispersion relations for the relativistic NS theory (left panel) 
and the CDR fluid dynamics (right panel) at the rest frame \cite{dkkm3}. 
The relativistic NS theory has two solutions, while the CDR fluid dynamics has three solutions.}
  \label{fig3}
\end{figure}

Then, are relativistic fluids Newtonian or non-Newtonian ?
To answer this question, we investigate the stability of these theories 
\cite{dkkm3,pu}.
Let us introduce a perturbation $\sim e^{i\omega t - ikx}$ around the 
hydrostatic equilibrium for the 1+1 dimensional system,
\begin{eqnarray}
\varepsilon = \varepsilon_0 + \delta \varepsilon~e^{i\omega t - ikx},~~ 
\Pi = \delta \Pi~e^{i\omega t - ikx}, ~~
\Xi = \delta \Xi~e^{i\omega t - ikx}, 
\end{eqnarray}
where $\varepsilon_0 = {\rm const}$ and the fluid velocity is 
parameterized with $\Xi$ as $u^\mu =(\cosh \Xi,\sinh \Xi) $.
Then the linearized CDR fluid dynamics is summarized as 
\begin{equation}
A X = 0,
\end{equation}
where 
\begin{eqnarray}
X &=& (\delta \varepsilon, \delta \Xi, \delta \Pi), \\
A &=& 
\left(
\begin{array}{ccc}
i\omega & -ik (\varepsilon + P) & 0 \\
-ik c^2_s & i\omega (\varepsilon + P) & -ik \\
0 & ik \zeta & 1+i\omega \tau_\Pi
\end{array}
\right),
\end{eqnarray}
where $P_0 = P(\varepsilon_0)$, and 
$c_s$ is the velocity of sound. In the following calculation, 
we consider a massless ideal gas, where $\varepsilon_0 = 3 P_0$ and 
$c^2_s = dP/d\varepsilon = 1/3$. 
Note that the result of the relativistic NS equation 
is reproduced by taking $\tau_\Pi = 0$.

The dispersion relation is obtained from ${\rm det}~A = 0$.
When $\tau_\Pi \neq 0$, we have one non-propagating mode and two propagating modes.
From the propagating modes, the group velocity is calculated as 
\begin{equation}
v_G = \frac{\partial {\rm Re} \omega}{\partial k} 
\approx \sqrt{c^2_s + \frac{\zeta}{\tau_\Pi (\varepsilon_0 + P_0)}}.
\end{equation}
One can see that the group velocity diverges in the vanishing $\tau_\Pi$ limit. 
That is, the relativistic NS theory contains infinite velocity propagations.

In order to study the stability, the imaginary part of the 
dispersion relations is shown in Fig. \ref{fig3}.
We used $\zeta/s = 0.1$ and $(\zeta/s)/(\tau_\Pi/\beta) = 1/6$ where 
$s$ is the entropy density and $\beta$ is the inverse of temperature.
In this parameter set, the group velocity is $v_G = 1/\sqrt{2}$, which is a causal 
parameter set because the speed is slower the speed of light.
The relativistic NS equation (left panel) has two solutions and 
the CDR fluid dynamics (right panel) has three solutions.
All the imaginary parts are positive and both theories are stable.
This result was already known in Ref. \cite{his}.

\begin{figure}
  \includegraphics[height=.25\textheight]{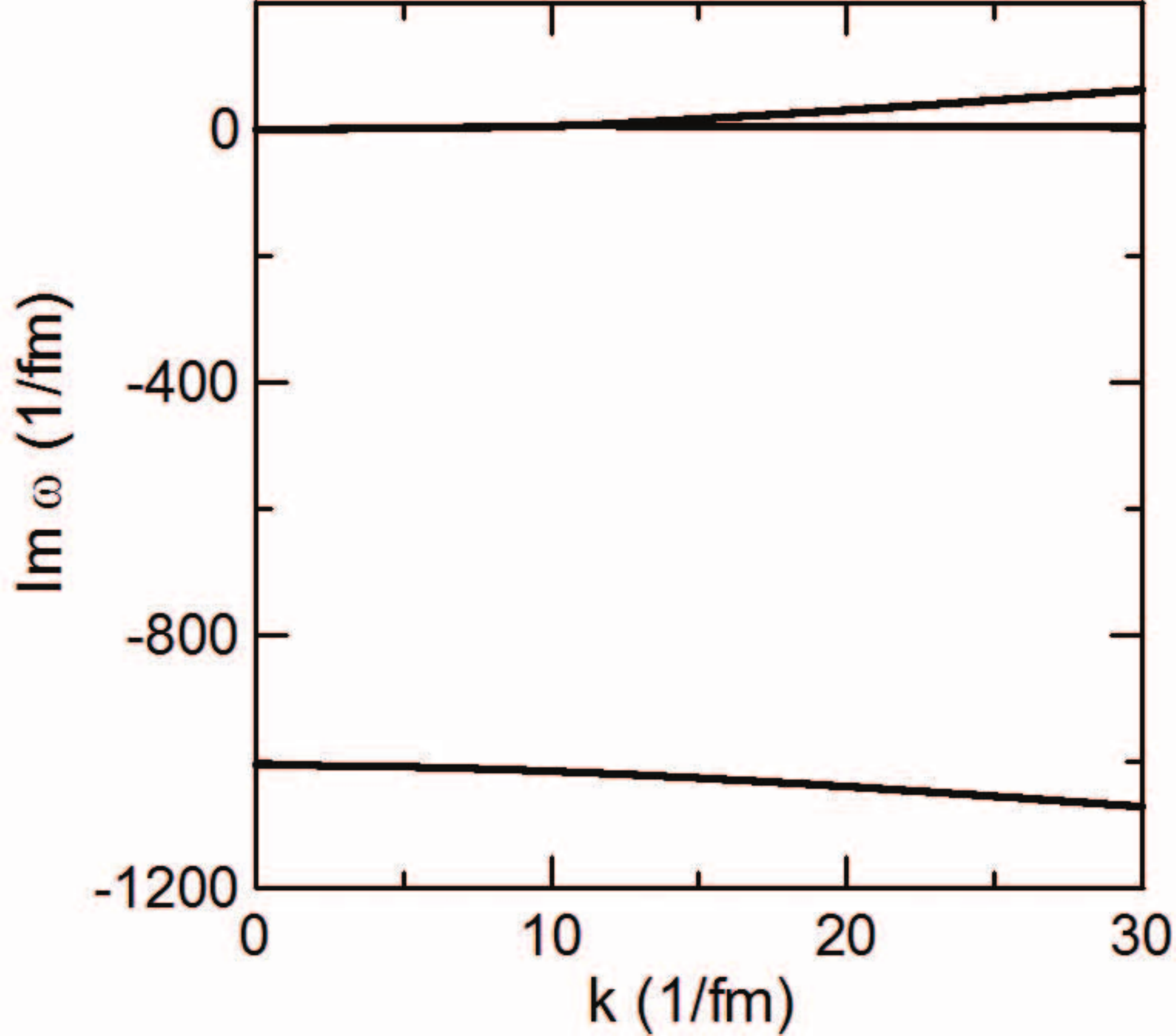}
  \includegraphics[height=.25\textheight]{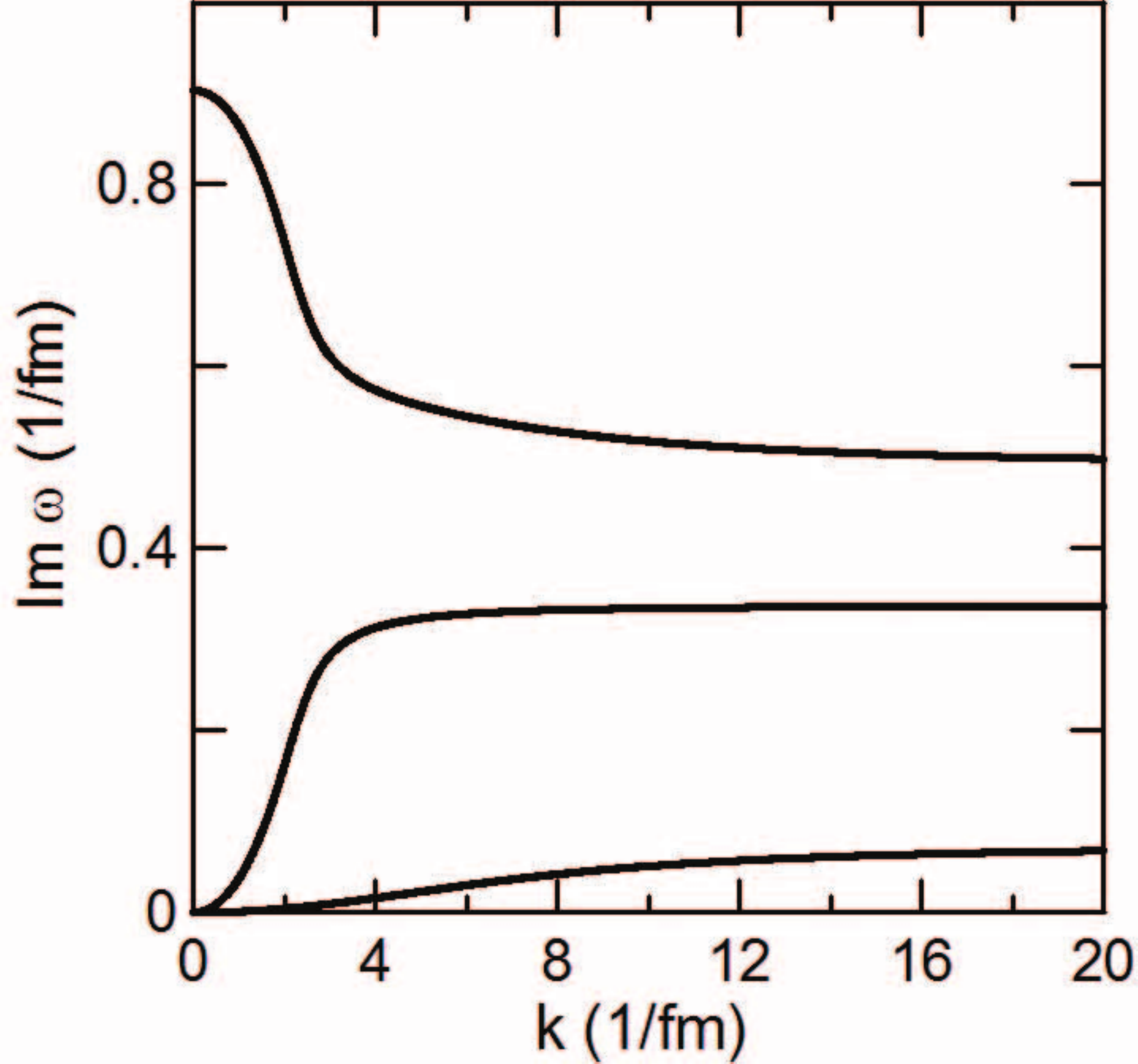}
  \caption{The imaginary part of the dispersion relations for the relativistic NS theory (left panel) 
and the CDR fluid dynamics (right panel) at the rest frame \cite{dkkm3}. In this case, both theories have three 
solutions. One of the solution of the relativistic NS theory is negative.}
  \label{fig4}
\end{figure}

If the theories are consistent with the relativistic kinematics, 
the nature of stability should not be changed by the Lorentz transform.
To see this, we study the stability from a Lorentz boosted frame 
with a boost velocity of $V=0.1$. 
As is shown in Fig. \ref{fig4}, one of the imaginary parts of the relativistic 
NS equation (left panel) becomes negative.
On the other hand, the imaginary parts of the CDR fluid dynamics (right panel) 
are always positive.
That is, the relativistic NS equation is inconsistent 
and inadequate as the theory to describe relativistic dynamics.

These results suggest that the violation of causality and instability are correlated.
To confirm this, 
we study the stability of the CDR fluid dynamics with an acausal parameter set, 
$(\zeta/s)/(\tau_\Pi/\beta) = 1$, where the group velocity exceeds the speed of light, $v_G = \sqrt4/3$.
The result is given in Ref. \cite{dkkm4}.
Again, one of the imaginary part becomes negative in a Lorentz boosted frame, although 
all the imaginary parts are positive in the rest frame.
That is, the violation of causality and instability is intimately related, and 
it is concluded that relativistic fluids are non-Newtonian. 
The results are summarized in table \ref{tab:b}.
So far, we have discussed the bulk viscous pressure. The same result is obtained 
even for the shear viscous tensor \cite{pu}.

\begin{center}
\begin{table}
\begin{tabular}{c c c c}
\hline
        &  RNS    & CDR (acausal parameter) & CDR(causal parameter) \\
\hline
Rest frame & stable & stable & stable \\
Boosted frame & unstable & unstable & stable \\
\hline
\end{tabular}
\caption{Relation between causality and stability \cite{dkkm3}}
\label{tab:b}
\end{table}
\end{center}

\section{Transport coefficients of CDR fluid dynamics}

In fluid dynamics, 
transport coefficients are inputs which should be calculated 
from the underlying microscopic dynamics.
In classical and non-relativistic NS fluids, it is known that 
the shear viscosity, for example, shows the following density dependence 
\footnote{So far, the coefficient $\eta_2$ has not measured experimentally. 
It will mean that $\rho$ is not a good expansion parameter of $\eta$ \cite{gar}. }\cite{dorfman}, 
\begin{equation}
\eta = \eta_0 + \eta_1 \rho + \eta_2 \rho^2 \ln \rho + \cdots. \label{eta_den}
\end{equation}
The first term $\eta_0$ can be calculated from two different approaches: 
the Chapman-Enskog expansion of the Boltzmann equation 
and the Green-Kubo-Nakano (GKN) formula.
It is known that the both results are consistent.
On the other hand, the higher order coefficients $\eta_1$ and $\eta_2$ 
are not calculated from the Boltzmann 
equation and we should use the GKN formula \footnote{ The Bogoliubov-Choh-Uhlenbeck equation 
should be used for calculating the higher order terms in the kinetic approach.}.

However, we cannot use the GKN formula to estimate the transport coefficients of 
the CDR fluid dynamics, because the GKN formula is derived by assuming 
the fluid to be Newtonian. 
Thus we have to derive a new formula to calculate the transport coefficients of 
the CDR fluid dynamics.

The formula is derived by using the projection operator method \cite{koide,dhkr}. 
The results are summarized as 
\begin{eqnarray}
&& \frac{\eta}{\beta (\varepsilon + P)} 
=\frac{\eta_{GKN}}{\beta^2 \int d^3 {\bf x} (\hat{T}^{0x}({\bf x}),\hat{T}^{0x}({\bf 0}))}, 
~~
\frac{\tau_{\pi}}{\beta}
=
\frac{\eta_{GKN}}{\beta^2 \int d^3 {\bf x} (\hat{T}^{yx}({\bf x}),\hat{T}^{yx}({\bf 0}))}, 
\label{taueta}\\
&& \frac{\zeta}{\beta (\varepsilon + P)} 
= 
\frac{\zeta_{GKN}}{\beta^2 \int d^3 {\bf x} (\hat{T}^{0x}({\bf x}),\hat{T}^{0x}({\bf 0}))}, 
~~
\frac{\tau_{\Pi}}{\beta}
= 
\frac{\zeta_{GKN}}{\beta^2 \int d^3 {\bf x} (\delta \hat{\Pi}({\bf x}),\delta \hat{\Pi}({\bf 0}))}, 
\label{tauzeta}
\end{eqnarray}
where $\hat{~}$ denotes operator, 
$\hat{\Pi} = \sum_{i=1}^3 \hat{T}^{ii}/3 - c^2_s \hat{T}^{00}$ and 
$\delta \hat{A} = \hat{A} - {\rm Tr}[\rho_{eq}\hat{A}]$ with the equilibrium density matrix 
$\rho_{eq}$.
Here $\eta_{GKN}$ and $\zeta_{GKN}$ are the shear and bulk viscosities of Newtonian fluids 
which are calculated using the GKN formula (more exactly, the Zubarev method). 
One can see that the new transport coefficients are still 
calculated from the GKN formula with the normalization factors given by the static correlation functions.

\begin{figure}
  \includegraphics[height=.25\textheight]{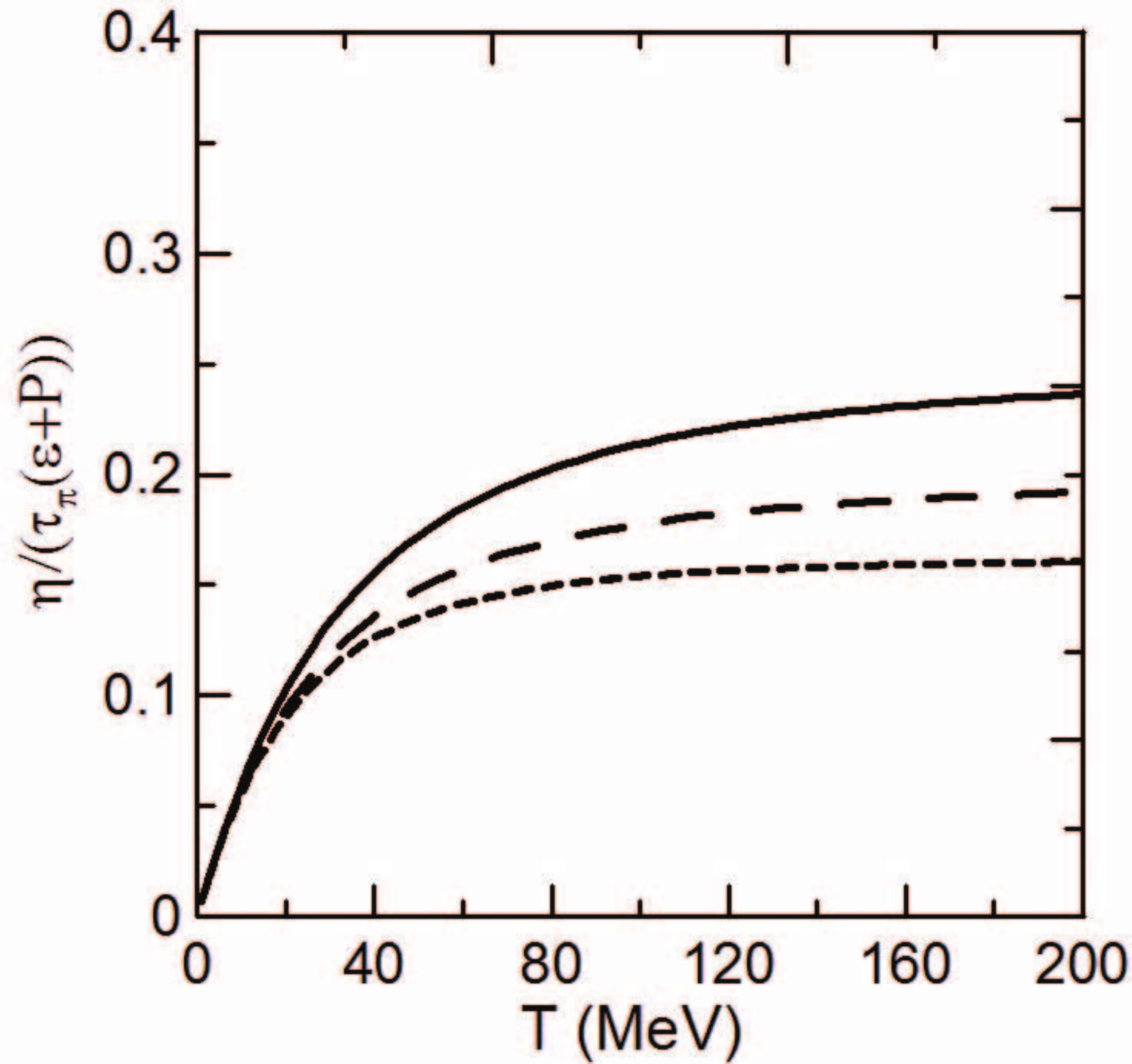}
  \includegraphics[height=.25\textheight]{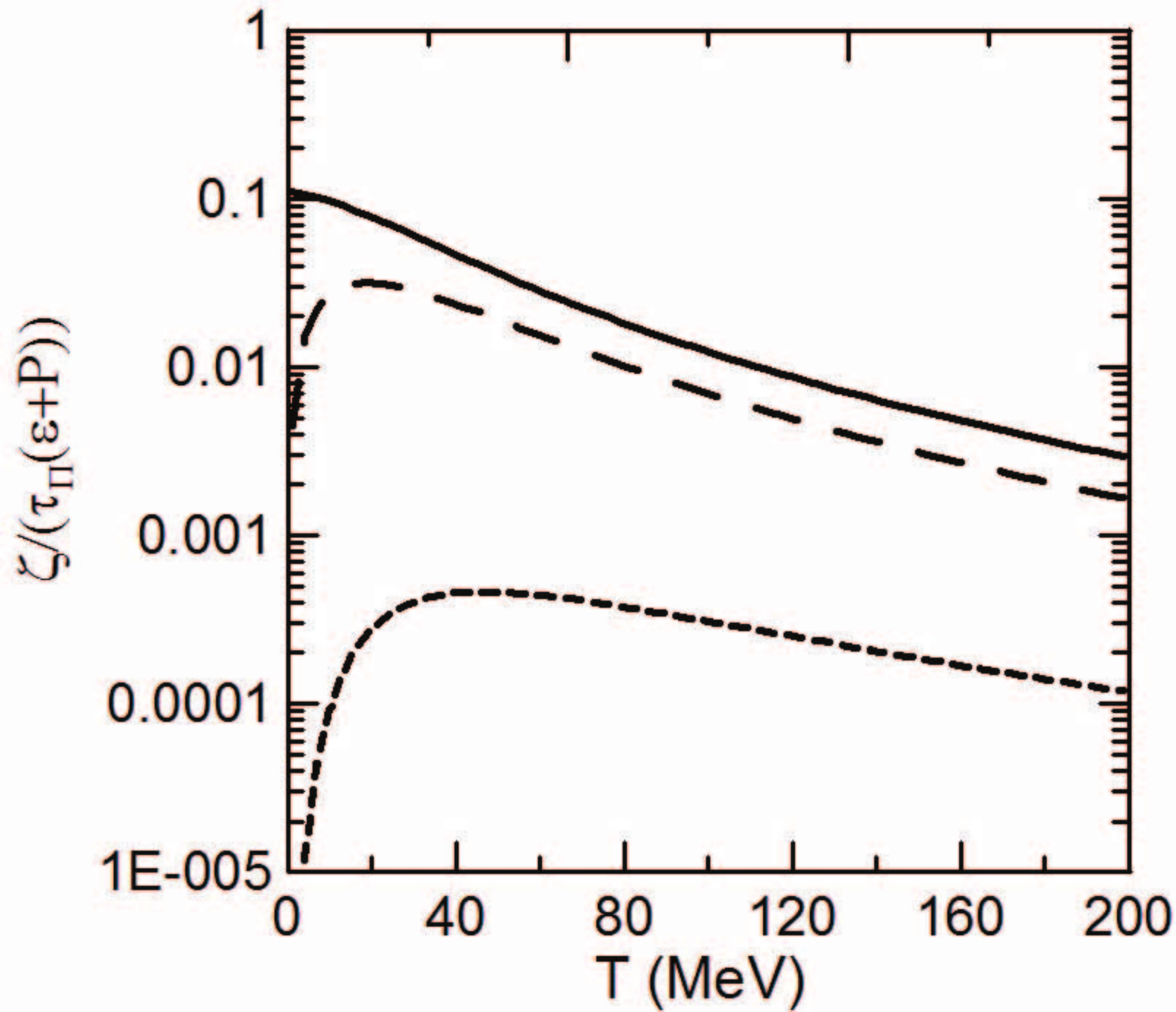}
  \caption{The temperature 
dependences of $\eta/(\tau_\pi(\varepsilon+P))$ (left panel) 
and $\zeta/(\tau_\Pi(\varepsilon+P))$ (right panel)
for pions, $m=140$ MeV \cite{dhkr}. 
The solid, dashed, and dotted lines 
represent the results of the microscopic formula, 
the new kinetic calculation, and the IS calculation, respectively.}
  \label{fig5}
\end{figure}

The same coefficients can be calculated from the Boltzmann equation 
with Grad's moment method \cite{is,gab}.
Now we compare the new formula with the Boltzmann equation.
For this purpose, 
we calculate the quantities $\eta/(\tau_{\pi}(\varepsilon + P))$ 
and $\zeta/(\tau_\Pi (\varepsilon + P))$, because these  
are independent of the 
choice of the collision term of the Boltzmann equation. 
The behaviors of $\eta/(\tau_{\pi}(\varepsilon + P))$ 
and $\zeta/(\tau_\Pi (\varepsilon + P))$ are shown on the left and right hand sides 
of Fig. \ref{fig5}, respectively.
The results from the new formula are shown by the solid lines.
Note that, as was discussed in Eq. (\ref{eta_den}), the Boltzmann equation can be consistent with 
the microscopic formula only in the dilute gas limit.
Thus we calculate the ratios in the leading order perturbative approximation.

So far, two different results for the same ratios are known from the Boltzmann equation; 
one is the well-known result obtained by Israel and Stewart (IS) \cite{is}, and the other 
the result obtained recently by Denicol et al. (DKR) \cite{gab} 
\footnote{ The same ambiguity of Grad's moment method is recently discussed even for non-relativistic cases 
\cite{garcia2}.}.
The former and latter are plotted by the dotted and dashed lines, respectively.
The DKR results predict larger ratios than those of IS, but are still smaller than the 
results of the microscopic formula.

This difference comes from the effect of quantum fluctuations.
In order to incorporate the quantum effect in the Boltzmann equation, 
the collision term is modified.
The ratios are, however, independent of the collision term and hence 
quantum corrections are not included.
Thus, to compare the two results, the effect of quantum fluctuations should be 
neglected.
Then we find that the solid lines agree with the dashed lines \cite{dhkr}.
As is shown in Fig. \ref{fig5}, the effect of these fluctuations is quantitatively large and 
cannot be ignored even in the high temperature limit.

\section{Summary}

We discussed the infinite propagation speed of the diffusion equation and 
introduced the Maxwell-Cattaneo-Vernotte (MCV) equation to solve this problem.
The drawback and advantage of the diffusion and MCV equations are summarized in table \ref{tab:a}.
The MCV equation should be modified when there exists a macroscopic flow.

By using this modified MCV equation, we derived a relativistic fluid-dynamical model called the causal 
dissipative relativistic (CDR) fluid dynamics. On the other hand, another relativistic fluid model 
was obtained by the relativistic generalization of the Navier-Stokes (NS) theory.
To see which theory is more adequate, the stability of the theories was studied 
with linear analysis. Then we found that the violation of causality and instability are 
intimately related and the theory becomes unstable if it contains acausal propagations.
The relativistic NS theory contains such propagation,  
and hence is unstable, while the CDR fluid dynamics is causal and stable.
In this sense, all relativistic fluids must be non-Newtonian.

Finally, we discussed the calculation of transport coefficients of the CDR fluid dynamics.
Because of their non-Newtonian nature, the Green-Kubo-Nakano (GKN) formula is not applicable.
The formulae for the CDR fluid dynamics are shown in Eqs. (\ref{taueta}) and (\ref{tauzeta}).
These formulae are consistent with the results of the Boltzmann equation with Grad's moment method.


\vspace{2cm}

This work was supported by CNPq, FAPERJ and the Helmholtz International
Center for FAIR within the framework of the LOEWE
program launched by the State of Hesse.






\end{document}